\documentclass{SCGEcontent}
\usepackage{titletoc}
\begin{document}

\begin{picture}(0,0){\rm
\put(0,-20){\makebox[160truemm][l]{\bf {\sanhao\raisebox{2pt}{.}}
Invited Review  {\sanhao\raisebox{1.5pt}{.}}}}}
\put(0,-34){\jiuwuhao {\textcolor[rgb]{0.5,0.5,0.5}{\sf Special Topic: the Next Detectors for Gravitational Wave Astronomy
}}}
\end{picture}

\def\bm{\boldsymbol}

\def\dl{\displaystyle}
\def\du{\end{document}}
\def\d{{\rm d}}
\def\e{{\rm e}}
\def\r{{\bm r}}
\def\P{{\bm P}}
\def\A{{\bm A}}
\def\k{{\bm k}}
\def\Q{{\bm Q}}
\def\pi{{\uppi}}
\def\cp#1{\mathbf{#1}}

\Year{2015} %
\Month{December} %
\Vol{X} 
\No{X} 
\BeginPage{1} 
\EndPage{11} 
\AuthorMark{{\rm Lee H M}, et al.}  
\AuthorMarkCite{{\rm Lee H M, Le B E-O, Du Z H}, et al.} 
\DOI{10.1007/s11433-015-5740-1} 
\ArtNo{X}

\title{Gravitational wave astrophysics, data analysis and multimessenger astronomy\footnotemark[2]\footnotetext[2]{Sect. 1 is contributed by LEE Hyung Mok (email: hmlee@snu.ac.kr); sect. 2 is contributed by LE~BIGOT Eric-Olivier, DU ZhiHui, LIN ZhangXi (corresponding
author, LE~BIGOT Eric-Olivier, email: eric.lebigot@normalesup.org); sect. 3 is contributed by DU ZhiHui, GUO XiangYu, WEN LinQing (corresponding author, DU ZhiHui, email: duzh@tsinghua.edu.cn); sect. 4 is contributed by
PHUKON Khun Sang, PANDEY Vihan, BOSE Sukanta (corresponding author, BOSE Sukanta, email: sukanta@wsu.edu); sect. 5 is contributed by FAN Xi-Long, HENDRY Martin (corresponding author, FAN Xi-Long, email: fanxilong@outlook.com)\\
$^{\dag\dag}$Permanent address: Laboratoire AstroParticule et Cosmologie (APC), Universit
  Paris Diderot-Paris 7, B$\rm c\hspace*{-1mm}/$timent Condorcet, Case 7020, 75205 Paris
  Cedex 13, France}}

\author[1]{LEE Hyung Mok}{}
\author[2\dag\dag]{LE~BIGOT Eric-Olivier}{}
\author[3]{DU ZhiHui}{}
\author[3]{LIN ZhangXi}{}
\author[3]{GUO XiangYu}{}
\author[4]{\vspace*{1.3mm}\\WEN LinQing}{}
\author[5]{PHUKON Khun Sang}{}
\author[6]{PANDEY Vihan}{}
\author[6,7]{BOSE Sukanta}{}
\author[8]{\vspace*{1.3mm}\\FAN Xi-Long}{}
\author[9]{HENDRY Martin}{}

\address[{\rm1}]{Department of Physics and Astronomy, Seoul National University, Seoul 08826, Korea;}
\address[{\rm2}]{Research Institute of Information Technology,
 Tsinghua   National Laboratory for Information Science and Technology, \\Tsinghua   University, Beijing 100084, China;}
\address[{\rm3}]{Tsinghua National Laboratory for Information Science and Technology,  Department of Computer Science and Technology, \\Tsinghua University, Beijing 100084, China;}
\address[{\rm4}]{School of Physics, The University of Western Australia, M468, 35 Stirling Hwy, Crawley, WA 6009, Australia;}

\address[{\rm5}]{Department of Physics, IIT Kanpur, Kanpur 208 016, India;}
\address[{\rm6}]{Inter-University Centre for Astronomy and Astrophysics, Post Bag 4, Ganeshkhind, Pune 411 007, India;}
\address[{\rm7}]{Department of Physics \& Astronomy, Washington State University,
1245 Webster, Pullman, WA 99164-2814, USA;}

\address[{\rm8}]{School of Physics and Electronics Information, Hubei University of Education, Wuhan 430205, China;}
\address[{\rm9}]{SUPA, School of Physics and Astronomy, University of Glasgow, Glasgow G12 8QQ, UK}

\maketitle \vspace{-3.5mm}{\footnotesize\begin{center} Received September 24, 2015; accepted September 28, 2015
\end{center}}\vspace*{-5mm}

\begin{center}
\rule{16.5cm}{0.4pt}
\parbox{16.5cm}
{\begin{abstract}This paper reviews gravitational wave sources and their detection. One of the most exciting potential sources of gravitational waves are coalescing binary black hole systems. They can occur on all mass scales and be formed in numerous ways, many of which are not understood.  They are generally invisible in electromagnetic waves, and they provide opportunities for deep investigation of Einstein's general theory of relativity. Sect. 1 of this paper considers ways that binary black holes can be created in the universe, and includes the prediction that binary black hole coalescence events are likely to be the first gravitational wave sources to be detected. The next parts of this paper address the detection of chirp waveforms from coalescence events in noisy data. Such analysis is computationally intensive. Sect. 2 reviews a new and powerful method of signal detection based on the GPU-implemented summed parallel infinite impulse response filters. Such filters are intrinsically real time alorithms, that can be used to rapidly detect and localise signals. Sect. 3 of the paper reviews the use of GPU processors for rapid searching for gravitational wave bursts that can arise from black hole births and coalescences. In sect. 4 the use of GPU processors to enable fast efficient statistical significance testing of gravitational wave event candidates is reviewed. Sect. 5 of this paper addresses the method of multimessenger astronomy where the discovery of electromagnetic counterparts of gravitational wave events can be used to identify sources, understand their nature and obtain much greater science outcomes from each identified event.
\end{abstract}}
\end{center}\vspace*{-0.6cm}

\begin{center}
\parbox{16.5cm}
{\bf\jiuhao gravitational waves, data analysis, multimessenger}
\end{center}

\begin{center}
{}
\Cit{Lee H M, Le Bigot E-O, Du Z H, et al. Gravitational wave astrophysics, data analysis and multimessenger astronomy. Sci China-Phys Mech Astron, 2015, 58: 120403,
 doi: 10.1007/s11433-015-5740-1}
\end{center}

\textwidth=178truemm \textheight=236truemm

\wuhao\vspace*{1.5mm}
\tableofcontents
\vspace*{5mm}
\begin{multicols}{2}

\renewcommand{\baselinestretch}{1.08} \baselineskip 12.2pt\parindent=10.8pt

\renewcommand{\thefootnote}

\section{Observation of gravitational waves as a deep probe of general relativity}
\emph{Binaries composed of compact stars such as neutron stars and black holes are prime targets for the ground detectors such as LIGO and Virgo. Such binaries are thought to have formed through the evolution of binary stars with massive stars. However, dense stellar systems such as globular clusters and nuclei clusters in the central parts of galaxies can also form compact binaries through dynamical processes. Recent studies showed that a few tens of black hole merger events due to dynamically formed binaries can be seen every year by the advanced detectors.}

\vspace*{-1mm}
\subsection{The rate of dynamical formation of compact black hole binary systems in the universe}
\vspace*{-1mm}

Several binary pulsars composed of two~ neutron~ stars~ have

\noindent been discovered so far. Based on the five systems whose merger time scales are shorter than
the Hubble time,  the expected rate of mergers of neutron star
binaries in our Galaxy has been estimated \cite{1Kim}. Assuming that the density of Milky Way
equivalent galaxy of 0.0116 Mpc$^{-3}$, the expected merger rate for the advanced detectors has been
estimated to be $c$. That gives the expected detection rate
of 8$_{-5}^{+10}$ per year.

The estimation of black hole binary merger rate is much more uncertain because no such binaries
have been discovered so far as the black holes do not emit electromagnetic radiation. Number of black
hole binaries (BHB) and their orbital parameters can be only estimated by the population
synthesis models.
Furthermore, the the mass of the black holes formed by the stellar evolution is not well known
either. The range of the BHB merger depends on the masses of the black holes in the binary,
leading to further uncertainty in the estimation of the
event rates for the advanced detectors. Taking such
uncertainties into account, the estimated event rates of black hole binaries  lies between 0.4 to 1000 per year with median value of 20 ${\rm yr^{-1}}$ \cite{2Abadie}.

Some binaries can be composed of a neutron star and a black hole. The merger rate of such binaries to
be detected by advanced detectors is estimated to lie between 0.2 to 300 ${\rm yr^{-1}}$ with the median
value of about 10 ${\rm yr^{-1}}$ \cite{2Abadie}. Either  binary neutron stars (BNS) or NS-BH merger will also
accompany with
electromagnetic radiation. The short gamma-ray bursts (SGRB) are thought to be due to merger
involving at least one neutron star.
However, the rate of SGRB is much smaller than the estimated merger rates of BNS or NS-BH
binaries. Such a discrepancy can be understood by the fact that the GRBs are highly beamed
events so that the GW and SGRB will be seen at the
same time only when the orbital planes of the merging binaries are almost perpendicular to the
line of sight. Nearly isotropic radiation is expected in the form of after glows
following the GW event, but the luminosity of such after glow is not very high.

When making such estimates, most of the binary systems are assumed to be originated by the
evolution of binary stars composed of massive stars. However, the binaries can also be formed
in dense stellar environment through dynamical processes such as tidal capture \cite{3Press,4Lee}, gravitational
radiation capture \cite{5Peters} or three-body processes \cite{6Goodman}. Tidal capture requires at least
one normal or giant star but the three-body processes does not. The central parts of globular clusters
or nuclei star clusters in the central parts of galaxies can provide suitable environments for the
the dynamical formation of compact binaries that can become GW sources.
The GWs coming from dynamically formed binaries can be additional sources to the
ones produced through the evolution of primordial binaries.

\subsubsection{Dynamical evolution of dense stellar systems and binary formation}
Dynamical evolution of dense stellar systems has been studied in great detail with various means.
The course of evolution depends on the existence of the central black hole of significant mass. The globular clusters are generally regarded as the systems without
a central black hole while the central parts of the galaxies are known to harbor
supermassive black holes (SMBH). There have been some reports on the existence of the
intermediate mass black holes (IMBH) in the center of the globular clusters \cite{7Farrell}, but the dynamical effect
of IMBHs to the globular cluster may not be significant. We simply assume no IMBH in the globular
clusters. We consider a central black hole only for the nuclear clusters in the galaxies.

It has been known for a long time that the self-gravitating systems (without a central black hole)
evolve in such a way that the central
parts become denser while the outer parts expands very slowly. The presence of
the mass function leads to the mass segregation because of the tendency to reach equipartition
among components with different masses.

Consider spherical systems composed of  two components of $m_1$ and $m_2 (> m_1)$.
The massive component   settle down to the central parts by dynamical friction on a time scale of
$(m_1/m_2 )~ t_{\rm rel}$ where $t_{\rm rel}$ is the relaxation
time \cite{8Lee2}. Note that the single-component cluster undergoes core collapse in several relaxation
time scales. Once the central parts
are dominated by the massive component, the subsystem of $m_2$ undergoes core collapse
on the subsytem's  own relaxation time scale, which is shorter than average relaxation time
of the entire system.

As the density of the central parts becomes higher binaries can be formed by either tidal capture
or three-body processes. Since we will be assuming neutron stars or black holes as high mass
components, we will only consider three-body processes and gravitational radiation capture.

The formation rate of three-body binary with sufficiently large binding energy for survival per volume for stars with mass $m$ and one-dimensional
velocity dispersion $\sigma$ can be expressed as \cite{6Goodman}:
\begin{equation}
{\d n_{\rm B}\over \d t} \sim G^5 m^5 n^3 \sigma^{-9},
\end{equation}
where $G$ is the the Newton's constant. On the other hand, gravitational radiation capture takes place
when two stars of mass
$m$ experiences encounters with the largest pricenter distance of
\begin{equation}
r_{\rm p,max} = \left[ {85\pi {\sqrt 2} \over 12} {G^{7/2}\over c^5}  {2^{3/2} m^{7/2}\over v_\infty^2}\right]^{2/7},
\end{equation}
where $v_\infty$ is the relative velocity of two stars at infinity. The rate of capture per volume
can then be obtained by
applying the capture cross section $\Sigma_{\rm cap}$
\begin{equation}
{\d n_{\rm cap}\over \d t} = {1\over 2} n^2  \langle \Sigma_{\rm cap} v_\infty \rangle \approx 9.6 {Gm^2\over c^{10/7} }
\sigma^{-11/7},
\end{equation}
where we used the following capture cross section \cite{9Quinlan}:
\begin{equation}
\Sigma_{\rm cap} = 2\pi \left({85\pi\over 6\sqrt 2}\right)^{2/7} {G^2 w^{10/7} m^2
\over v^{10/7} v_\infty^{18/7}}.
\end{equation}
Assuming that the binaies are formed in the core of the stellar systems, the ratio between
the formation rates of binary
formation via three-body processes and the gravitational radiation capture can be
expressed as \cite{8Lee2,9Quinlan}:
\begin{equation}
{\d n_{3b}/\d t\over \d n_{\rm cap}/\d t} \approx 40 {1 \over N_{\rm c}^2} \left( {\sigma\over c}\right)^{10/7},
\end{equation}
where $N_{\rm c}$ is the number of stars in the core, which varies significantly through the dynamical
evolution but reaches a few tens during the core-collapse. Therefore, in most circumstances without
a central black hole the  three-body processes
are much more efficient than gravitational radiation capture.

An exception would be the vicinity of
a supermassive black hole in the galactic nuclei as the density and velocity dispersion can become
very high.
If there is a central black hole, the stellar system does not form a flat core, but cuspy distribution of
$n(r) \propto r^{-7/4}$, where $n(r)$ is the number density of stars as a function of radial
distance $r$ from the SMBH is expected \cite{11Bahcall}, while the velocity dispersion will follow the
Keplerian, i.e., $\sigma(r) \propto r^{-1/2}$. Close to the central black hole, the density and velocity
dispersion can become very high and the gravitational radiation capture becomes more important
than three-body processes.

The binaries can be characterized by the semi-major axis and the eccentricity. The semi-major axis can be translated
into hardness as defined by
\begin{equation}
x\equiv {Gm/2a\over 3\sigma^2/2},
\end{equation}
where $a$ is the semi-major axis. The binaries formed by three-body processes typically have $x\approx 3$ but the hardness
can change as a result of the dynamical interaction with other stars. The distribution of the eccentricity is expected to
be that of thermal, i.e., $f(e) \d e \propto e ~\d e$.

The orbital characteristics of the binaries formed by gravitational radiation capture are
quite different from those of three-body ninaries.
Typical values of the semi-major axis and eccentricity of the binaries formed by gravitational radiation
capture are \cite{12Hong}
\begin{equation}
a\sim 0.08~ {\rm AU}  \left( M\over 10~ {\rm M_\odot}\right) \left(\sigma\over 75~ {\rm km/s}\right)^{-2},
\end{equation}
and
\begin{equation}
e\sim 1-10^{-4} \left({\sigma\over 75~ {\rm km/s}}\right)^{-2}.
\end{equation}
Therefore, the gravitational radiation capture leads to the formation of very eccentric binaries, but the
hardness is not high.
Since the gravitational radiation is very efficient for binaries with small pericenter distance,
the orbit shrinks very rapidly.
Such binaries go into merger in a time much shorter than typical time scales for dynamcal
interactions with
other stars.  Gravitationally captured binaries in the stellar systems with velocity dispersion of about
100 km/s are expected to merge
in a few hundreds of years or less \cite{11Bahcall}.

On the other hand, the binaries formed by three-body processes have moderate eccentricity
and the time to
become merger is extremely long. The merger time scale can be expressed as:
\begin{eqnarray}
\tau_{\rm merge} &\hspace*{-2.5mm} = &\hspace*{-2.5mm}g(e) {405\over 8192} {Gmc^5\over x^4 {\sqrt 3} \sigma^8}\nonumber \\
&\hspace*{-2.5mm} \approx &\hspace*{-2.5mm} 4\times 10^{18}~ {\rm
yrs} \left({m\over 10~ {\rm M}_\odot}\right) \left({3\over x}\right)^{-4} \left({\sigma\over 10~ {\rm km/s}}\right)^{-8},
\end{eqnarray}
where
\begin{equation}
g(e) = (1-e^2)^{7/2} \left(1+{73\over 24 }e^2 + {37\over 96}e^4\right)^{-1}.
\end{equation}
In order for the binaries merge in Hubble time, their hardness should be significantly increased.

The dense environment provides further evolution of the binary orbits. The binaries have larger
cross sections than the single stars for dynamical interactions since the semi-major
axis can be regarded
as the size of the system. The close encounters between binaries and singles leads to the formation
of temporary triple system and leaves a binary with higher hardness at the expense of
higher relative velocity
between binaries and the singles. Such a  `hardening' of the binaries continues until the
recoil velocity of the
binaries exceeds the escape speed from the cluster. If the potential of the cluster is sufficiently deep,
some binaries can go into merger before they get ejected.

\subsubsection{ Globular clusters}
There are about 150 known globular clusters in the Galaxy and a few  tens of clusters are estimated to be hidden in the
direction where we cannot observe because of high interstellar extinction by dust \cite{13Ashman} .
The  stellar density in the central parts can become $10^4$--$10^6~ {\rm pc}^{-3}$ so that the dynamical interaction among
stars can be quite frequent. Most of the globular clusters are as old as the Galaxy itself, implying that they are one of the very first objects in the Galaxy.

Current globular clusters are composed of only low mass stars since most of the high mass stars have been evolved off.
There are some indication of the remnant stars such as white dwarfs and neutron stars in the form of low-mass X-ray binaries (LMXB)
and millisecond pulsars. The existence of stellar mass black holes in the globular clusters has been suggested by recent
 observations \cite{14Strader}. Such remnant stars
 must have been formed in early life of the clusters through the evolution of high mass stars.

The NSs (and some BHs) are formed after the supernova explosion. Any deviation of spherical
symmetry of the explosion leads to the kick velocity of the remnant stars. In fact many new born
neutron stars are known to be moving at large velocity (200--500 km/s) relative to the progenitor
stars. Since the escape velocity from globular clusters is only about 50 km/s, most of the
NSs born in globular clusters would have escaped almost immediately. However,
many millisecond pulsars that are thought to be old NSs are present in globular clusters.
For example,
Terezan 5 alone has 33 millisecond pulsars. It is not clear how such NSs survived in globular clusters,
but we may assume that there are still large number of neutron stars in current globular
clusters since only some fraction of NSs can become millisecond pulsars.
The kick velocity of BHs is not well known, but  could be as high as those of NSs \cite{15Janka} . The presence
of NSs and BHs in globular cluster may
indicate at least some fraction, which remains to be uncertain,
of remnant stars are formed with low kick velocity.

Despite of such uncertainties, there have been several studies regarding the BHBs from
globular clusters.  $N$-body code has been used to study the dynamical evolution of
globular clusters by varying the BH fractions and  obtained BHB merger events to be observed
by the advanced detectors of 31$\pm$ 7 yr$^{-1}$ from globular clusters \cite{16Banerjee}. Downing et al. \cite{17Downing}
obtained
the detection rate of 15 to 29 yr$^{-1}$ based on the study with Monte Carlo. More recently,
Tanikawa \cite{18Tanikawa}  obtained the detection rate of 0.5--20 yr$^{-1}$ based on the
$N$-body simulations assuming BH mass function that has a peat at around 5 M$_\odot$.
More recent study  with careful statistical analysis of the $N$-body simulations with
stellar mass BH \cite{19Bae} described below also found that significant number of BH binary merger events
will be observed by the advanced detectors.

The course of the evolution of globular clusters depends on the initial condition which is not well
understood. The evolution of massive stars leads to the rapid loss of mass from the cluster. The stellar
population should have changed in a short time compared to the age in such a way that most of
the high mass stars quickly turned into remnant stars. Relatively lower mass stars evolve slowly, producing white dwarf stars.
The presence of the primordial binaries can make the dynamical evolution
more complex. However, one may simplify the globular clusters in the following way: since the
time scale for the stellar evolution is very sensitive function of stellar mass, the formation of black
holes and neutron stars takes place during the very early phase of the evolution, the globular
clusters are composed of ordinary stars whose mass is of order of 0.7 M$_\odot$, neutron stars, and
black holes. We may assume that the typical masses of neutron stars and black holes are 1.4 and 10
M$_\odot$, respectively.

\begin{figure}[H]
\centering
\includegraphics[scale=1]{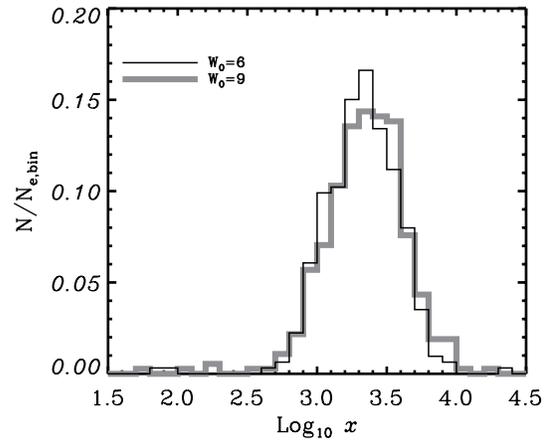}\vspace{-2mm}
\caption{The hardness distribution of ejected BH-BH binaries from the clusters. Initial models of the clusters are King models with
$W_0$ =6 and 9. The hardness distribution is found to be nearly independent of $W_0$ (Figure from ref. \cite{19Bae}).}
\label{fig1:BH}
\end{figure}

\begin{figure}[H]
\centering
\includegraphics[scale=1]{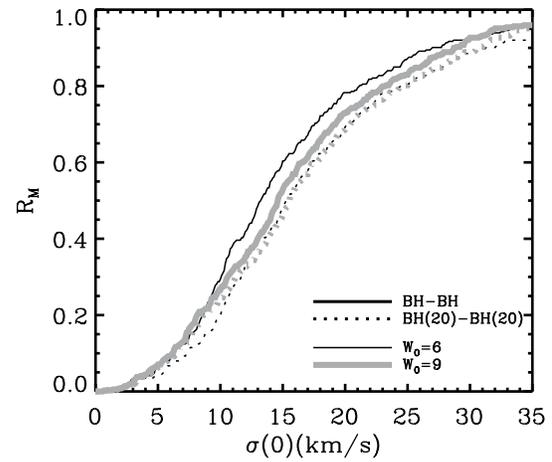}\vspace{-2mm}
\caption{The ratio of the total ejected binaries as a function of the central velocity dispersion of globular clusters for
models with
$W_0$ =6 and 9. Note that typical globular clusters have $3 < \sigma_0 < 15$ km/s. Dotted lines assume that the
mass of the black hole is 20 M$_\odot$ while solid lines assume 10  M$_\odot$ (Figure from ref. \cite{19Bae}).}
\label{fig2:ratio}
\end{figure}

Dynamical friction of black holes due to other stars quickly leads to the formation of subsystem
mainly composed of
black holes. Three-body binaries are efficiently formed under the physical conditions of the globular
clusters and they
undergo successive hardening through the encounters with other black holes. Most of the black holes
are evaporated
as a result of the close encounters between binaries and singles as described in the previous
subsection in very
short time scale (much smaller than the half-mass relaxation time). The $N$-body
simulation carried o\ showed that about 1/3 of the black holes are ejected in the form of
binaries and remaining 2/3 get ejected in the form of singles \cite{19Bae}. The distribution of hardness of the
escaping binaries obtained
from the $N$-body simulation is shown in Figure~\ref{fig1:BH}. The hardness distribution of
escaping binaries can be used to compute the
what fraction of the evaporated binaries will undergo gravitational wave merger within
Hubble time as a function of central
velocity dispersion ($\sigma_0$), as shown in Figure~\ref{fig2:ratio}. Because of the steep dependence of the
gravitational radiation merger
time scale on the semi-major axis, the fraction of merging binaries depends sensitively
on the velocity dispersion.
Typical globular cluster has $3 < \sigma_0 < 15$ km/s, and less than half of the escaped binaries will
merge in Hubble time.

The results obtained with $N$-body simulations can be used to estimate the merger rates of the BHB
systems for globular clusters in the Galaxy under
a number of assumptions. First, the number of black holes in the early phase of the cluster
evolution has been obtained by assuming  Kroupa mass\linebreak
\vspace{-7mm}

\noindent function \cite{20Kroupa}.
We have not considered any natal kick so that all black holes formed in the cluster are kept.
We then assumed that all black holes are evaporated
through binary formation and subsequent interactions. The number of evaporated binaries is
assumed to be 17\% of the total number of black holes.
We applied the fraction of merging binaries shown in Figure \ref{fig2:ratio} for each cluster with observed central
velocity dispersion. We have not
considered any possibility of variation of the velocity dispersion through dynamical evolution since the
it vevolves rather slowly.
We assigned expected number of merging binaries for 141 clusters whose parameters are compiled
by Gnedin et al. \cite{21Gnedin} and add them up to
obtain the merger rate within Hubble time in our Galaxy to derive
2.5 merger per globular clusters per Gyr.  The average number density of globular clusters
in the local universe is estimated to be about 2.9 Mpc$^{-3}$ \cite{22Portegies} and the horizon distances of
advanced LIGO-Virgo
network for BHB is 2187 Mpc. By multiplying the horizontal volume, we obtain the expected merger
rate of BHB for the advanced detectors of 15 yr$^{-1}$.
This is comparable to the estimated rate of about 20 yr$^{-1}$ from binary evolution of primordial
binaries \cite{2Abadie} and consistent with other estimates mentioned earlier \cite{16Banerjee,17Downing,18Tanikawa}.

Our estimate also could have large amount of uncertainties. First, large kick velocities at the birth of
BH could lead to substantial reduction of BHB formation.
Second, we assumed that all the BHs produced by supernova explosion underwent rapid evolution
and therefore evaporated from the cluster in much
shorter time than the current age. However, recent studies showed that the clusters with long
relaxation time might still possess large fraction
of the BHs.

On the other hand, there are various possibilities that our estimate is rather low. The total number of
globular clusters we observe today might not
represent the initial populations if large number of clusters have been completely disrupted due to
tidal evaporation or tidal shock \cite{23Gnedin}.
In fact the detailed studies with Fokker-Planck code showed that evaporation alone could have
destroyed more than 50\% of the initial population of globular clusters \cite{23Gnedin}. The disrupted clusters
should have left many compact BHBs since the rapid dynamical evolution leading to rapid
evaporation is a necessary condition for efficient BHB formation
Even for the present day globular clusters,  number of BHs in the the early phase may have been
significantly undersestimated. When we computed the number of BHs we used the present mass of
the clusters. However, the
clusters lose mass continuously, and some clusters have lost large fraction of the initial mass. If that
is the case, the number of BHs per cluster could
have been significantly larger. With all these possibilities, our estimates are rather conservative.
When all uncertainties are taken into account, the detection rate
could be as high as 60 yr$^{-1}$ .

BNS or binaries composed of a NS and a BH are also of great interest to the
advanced detectors as they can
be accompanied by the electromagnetic radiation in the form of SGRB or optical and radio
afterglows.
If large number of BNS are formed from globular clusters, they can explain the occurrence of the
SGRBs in the outer parts (or even outside) \cite{25Fong}
of the galaxies since the globular clusters reside in the halo and the binaries are ejected with
velocities higher than
the escape velocity from the cluster.

Neutron stars (NS) can also form a subsystem in the central parts after the black holes are
completely depleted.
Similar to the BHs, NSs can form binaries via three-body processes in the dense core and these
binaries undergo hardening process. However, the time scale for the formation of central subsystem
of NSs and subsequent evolution is much longer than the
case of BHs. Only clusters with short relaxation time would be able to eject most of the NSs via
binary-single interactions.
Our estimation of rate of BNS merger events with advanced detectors give very small numbers:
between 0.01 to 0.1
yr $^{-1}$, assuming 10\% for the retention of NSs in clusters. If the retention rate is higher, the
expected BNS merger
rate could become higher, but still much less than the that  from the evolution of primordial binaries.
We thus conclude that the
globular clusters could provide a only substantial number of BHB merger events but very little
BNS events.

The BHB events originating from globular clusters would be preferentially found from the outer parts
of the galaxies
in contrast to the disk populations. Since there are more globular clusters in
elliptical galaxies than in spiral galaxies,
the occurrence of the BHB merger events could be more easily seen near the vicinity of the giant
ellipticals that are more
concentrated in the center of cluster of galaxies since they have large number of
globular clusters. However, since the gravitational wave detectors have very poor
angular resolution it may be very difficult to see such an effect.

We also found that the dynamical formation of BH-NS binaries is
very unlikely because of the strong segregation of the
BH and NS. Inside the BH subsystem, the density of NS is much lower than that of BH. After the
depletion of the BHs, NSs cannot find BHs nearby.

\subsection{Nuclear star clusters}

Nuclear clusters are different from globular clusters in many aspects. The velocity dispersion of the
typical NCs is about 100 km/s which
is much larger than that of GCs. Most of the NCs are thought to have a supermassive black hole
(SMBH) in the center so that the stellar
density profile can be cuspy. The NC is also surrounded by the bulge and therefore the escape
velocity could be very large ($>1000$ km/s). There is no retention problem for the remnant stars
in the NCs.

The binary formation by gravitational radiation capture in NCs has been considered previously \cite{9Quinlan,26O'Leary}.
Similar to  the case of globular clusters,  individual mass of BHs is much larger than
that of ordinary stars. BHs inevitably experience dynamical friction that leads to the segregation
of BHs and lower mass components. However, the velocity dispersion would be the same
for all components regardless of the mass in the regime where the potential is
dominated by the SMBH.

O'Leary et al. \cite{26O'Leary}  used the density profiles of black holes in the vicinity of the SMBH obtained by
Hopman \& Alexander \cite{27Hopman}
and claimed that the rate of  the formation of BHBs via gravitational radiation capture would be
 5--20 yr$^{-1}$ within the horizon of
advanced detectors. As discussed earlier, such binaries immediate
merge since thepericenter distances are small and eccentricities are very large.
Therefore the formation rate is identical to  the merger rate.
These binaries could possess
significant amount of eccentricity when they enter into the advanced LIGO/Virgo band. Such an eccentric binaries
could be distinguished from the circular binaries that are formed by three-body processes or primordial binaries.

Recently, Hong \& Lee \cite{12Hong} visited this problem with careful $N$-body simulations.
The NCs are modeled by evolving
a spherical systems embedded by a surrounding potential well represented by a Plummer potential.
The central
SMBH was also taken into account by growing the mass of the~ BH adia-

\begin{figure}[H]
\centering
\includegraphics[scale=1]{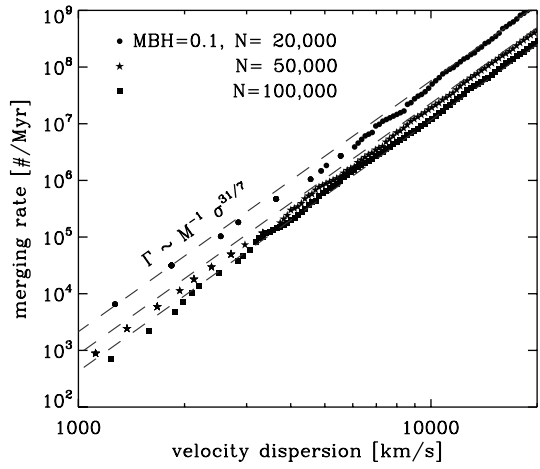}
\caption{Merger rates as a function of velocity dispersiopn for models with $M_{\rm SMBH}=0.1 M_{\rm tot}$ (Figure from ref. \cite{12Hong}). }
\label{fig3:merger}
\end{figure}

\noindent batically from small mass.
Eventually a stellar system with nearly Bahcall-Wolf cusp in the center and nearly isothermal
outer parts, mimicking the NC of the Galactic Center.

The rate of gravitational radiation capture is computed by using the criterion of eq. (2) and integrated over the
entire cluster using the density and velocity profiles of the $N$-body model. The integrated merger rate becomes

\begin{equation}
\Gamma_{\rm merger} \approx m^2 \cdot N\cdot {\tilde n} \cdot  \sigma_*^{-11/7} \sim M^{-1} \cdot \sigma_*^{31/7},
\end{equation}
where $\tilde n$, $\sigma_*$ and $M$ are mean number density, velocity dispersion of the ststem and the total mass of the
cluster, respectively. Virial theorem has been used to eliminate $\tilde n$ in favor of total mass and velocity dispersion.
Such a dependence has been verified by the numerical simulations as shown in Figure~\ref{fig3:merger}.
By extrapolating the simulation results to realistic parameters for NC, we obtain
\begin{equation}
\Gamma_{\rm merger} = C \left({M_{\rm SMBH}\over 3.5 \times 10^7~ {\rm M}_\odot}
\right)^{-1} \left({\sigma_*\over 75~ {\rm km/s}}\right)^{31/7},
\end{equation}
where the constant depends on the total mass of the cluster: $C=8.58 \times 10^{-5}~ {\rm Myr}^{-1}$ for $M_{\rm SMBH}/M=0.1$
and  $C=2.06 \times 10^{-4}~ {\rm Myr}^{-1}$ for  $M_{\rm SMBH}/M=0.2$.

The merger rate per volume is obtained by integrating the rate over the SMBH distribution, i.e.,
\begin{equation}
\mathcal{R}_{\rm mer}=\int^{M_{\rm u}}_{M_{\rm l}} \Gamma_{\rm merger}(M_{\rm SMBH})\frac{\d n_{\rm SMBH}}{\d M_{\rm SMBH}} \d M_{\rm SMBH},
\label{eq13}
 \end{equation}
where $M_{\rm u}$ and $M_{\rm l}$ are the upper and lower limits for integration, and $\Gamma_{\rm merger}$ is the merger rate per galaxy as a function of the mass of MBH.
The upper limit can be fixed to $M_{\rm SMBH}\sim 10^{7}~ {\rm M}_{\odot}$ by the time scale requirement: if the SMBH mass is too high, the relaxation time of the NC is so long that the stellar cusp
cannot form.
However, we still do not have the exact lower limit of the SMBH mass,
which is currently about $10^5~ {\rm M}_{\odot}$ from the observation of Barth et al. \cite{28Barth}.
With the uncertainty of the lower limit of the MBH mass from $10^3~ {\rm M}_{\odot}$ to $10^5~ {\rm M}_{\odot}$, eq. (\ref{eq13}) gives us the merger rate density
\begin{equation}
\mathcal{R}_{\rm merger}\approx (2-5) \Gamma_{\rm mer,MW}\xi_{30}~ {\rm Mpc}^{-3},
\end{equation}
where $\Gamma_{\rm mer,MW}$ is the merger rate for a Milky-Way-like galaxy, and
$\xi_{30}$ is the rescale factor for the variance of the number density of stars normalized by 30 (i.e., $1/3 \le \xi_{30} \le 3$).

Using this result, we can estimate the detection rates by advanced detectors: Hong \& Lee (2015) found that the expected deteciton rate of 0.05 to 4 yr$^{-1}$,
much smaller than previous estimates.  Thus we conclude that the eccentric binaries will be very rare with the advanced detectors.

\section{GPU acceleration for gravitational-wave burst searches}
\emph{We demonstrate that the current main gravitational-wave burst pipeline, Coherent WaveBurst, can be accelerated in some conditions by porting its CPU code to GPU. We emphasize the main principles that lead to the good GPU performance that we obtained. This work opens the possibility of implementing an improved burst detection algorithm that combine the signals from available gravitational-wave interferometers in a fully coherent way, thereby leading to a better rejection of noise.
}
\subsection{Introduction}

Some expected sources of gravitational waves produce waves with a
shape that is not easily modeled, like supernovae and gamma ray burst
events. In order to detect such typically short gravitational waves,
\emph{unmodeled} ``burst'' searches must be performed---this is one of
the main types of gravitational wave searches, along with searches for
Compact Binary Coalescences, and continuous and stochastic
gravitational waves. Such searches typically look for excess energy in
the signal measured by gravitational-wave interferometers.

Coherent WaveBurst is currently the main gravitational-wave burst
detection pipeline~\cite{klimenko2008}. It gathers together the signal
from multiple gravitational-wave interferometers (currently the two
LIGO ones, along with Virgo) \emph{coherently} (like some older
methods based on a maximum likelihood~\cite{pai2001} or on the
$F$-statistics~\cite{cutler2005}, and in non real-time Compact Binary Coalescence searches).

\begin{figure}[H]
\centering
\includegraphics[scale=1]{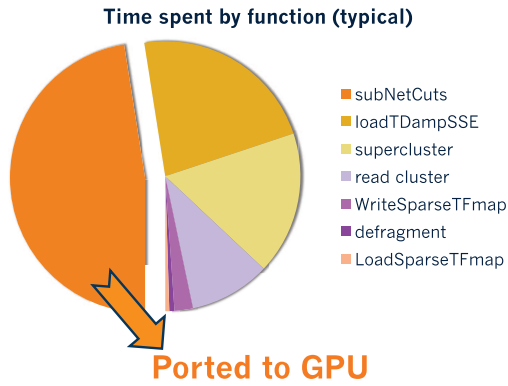}\vspace{-2mm}
\vspace*{1.5mm}
\caption{(Color online) Various parts of the Coherent WaveBurst
  pipeline, along with their fraction of the time spent during a
  typical data analysis run. The subNetCut function (called multiple
  times) takes the longest, and is therefore the one that was ported
  to GPU.}
\label{time_spent}
\end{figure}

Such a coherent burst search is, however, computationally
intensive. The production version of Coherent WaveBurst pipeline
therefore uses SSE CPU instructions, which process multiple data in
parallel. However, the level of parallelism of SSE instructions is
restricted to a few floating-point operations. We thus explored the
possibility of porting the most time consuming part of the pipeline,
the subNetCut function (see Figure~\ref{time_spent}), to the GPU.

This function essentially performs a loop over all the possible
locations of the gravitational-wave source in the sky, so as to find
its most likely position and, importantly, check that \emph{all} the
detectors coherently detect some signal (consistently with their
relative sensitivities). This function is thus instrumental in
eliminating spurious excess energy detected in the signal of only some
of the interferometers (glitches).

\subsection{Results}

\begin{figure}[H]
\centering
\includegraphics[scale=1]{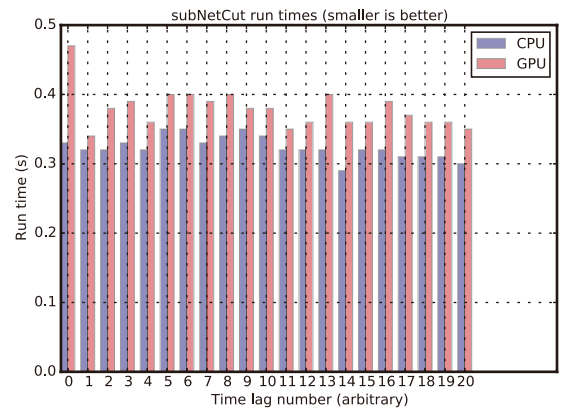}\vspace{-2mm}
\vspace*{1.5mm}
\caption{(Color online) Timing results for the function ported to GPU
  (subNetCut), with production parameters (setTDFilter~1 and
  HEALPIX~4). The GPU performs at about the same speed as the CPU.
}
\label{timing_CPU}
\end{figure}

\begin{figure}[H]
\centering
\includegraphics[scale=1]{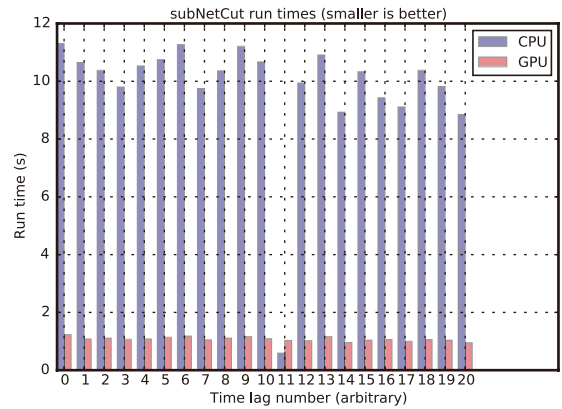}\vspace{-2mm}
\vspace*{1.5mm}
\caption{(Color online) Timing results for the function ported to GPU
  (subNetCut), with parameters for a more refined analysis
  (setTDFilter~4 and HEALPIX~7). The GPU is about 8 times faster than the CPU.
}
\label{timing_GPU}
\end{figure}

Our results show that such a partial GPU port of the Coherent
WaveBurst gravitational-wave detection pipeline is realistic and
potentially very useful: running with the current production
parameters of the pipeline takes essentially the same time (see
Figure~\ref{timing_CPU}), but running a realistic and more refined
analysis showed a 8-fold improvement in the computing speed of the
calculation (see Figure~\ref{timing_GPU}).

While the more refined analysis takes longer than the current
production analysis, it gives an estimate of the time that an
improved, ``fully coherent'' gravitational-wave detection algorithm
would take. Such an algorithm has not yet been implemented in Coherent
WaveBurst, but could replace the current subNetCut function, if it
runs sufficiently fast: the impact of such an improved algorithm on
the downstream computations of the pipeline is expected to be a
drastic reduction in data size and computing time. Our results
demonstrate that implementing a fully coherent algorithm in cWB is an
improvement that can be reasonably considered; this would improve the
sensitivity of the gravitational-wave burst detection.

\subsection{Porting to GPU: key principles}

An important principle behind GPU acceleration is naturally the need
to express the code to be ported a parallel calculations. Because of
synchronization constraints, it is also important that all the
calculations performed in parallel take about the~ same~
time---otherwise~ some~ GPU~ cores~ remain

\noindent idle. We thus had to completely
rewrite the structure of the CPU version (of the subNetCut function)
so as to achieve this level of parallelism.

Concretely, the CPU version uses SSE instructions that perform
calculations in parallel over four pixels of a time-frequency
transform of the gravitational-wave interferometer signal. The
rewritten GPU version performs instead a completely different parallel
calculation, over 4096 different potential gravitational-wave source
locations.

Another important principle of efficiently porting to current GPUs is
that data transfers must be performed between the CPU and the GPU
before calculations can be run on it. Such memory transfers can limit
the overall performance of the calculation. Part of the rewriting
effort thus also focused on minimizing the amount of data exchanged
between the CPU and the GPU (typically by defining where in the code
the CPU should hand over calculations to the GPU, and where the GPU
could stop its calculations and send the results back). We observed
substantial speed gains after this data optimization.

\subsection{Conclusion}

We have demonstrated the feasibility of porting to GPU a
computationally critical function of the Coherent WaveBurst
gravitational-wave detection pipeline. These results open the
possibility of implementing a more powerful algorithm for detecting
gravitational-wave signals coherently between multiple
interferometers, that would lead to much smaller output
(gravitational-wave trigger) files and smaller overall pipeline run
times.

\subsection{Appendix: technical details}

The timings reported here were obtained with the following hardware:

$\bullet$ GPU: NVIDIA GF100 (GeForce GTX 480)

\qquad $-$ Compute capability: 2.0

\qquad $-$ CUDA Cores: 480

\qquad $-$ Processor Clock: 1401 MHz

\qquad $-$ Memory Clock: 1848 MHz

\qquad $-$ Memory Size: 1536~MB

\qquad $-$ Memory Bandwidth: 177.4~GB/sec

\qquad $-$ Warp Size: 32

$\bullet$ CPU: Intel Core i3-2100 CPU, 3.10~GHz

\section{GPU accelerated SPIIR method}
\emph{Summed Parallel Infinite Impulse Response (SPIIR) filtering technique  is unique because it directly employs parallel computing idea into solving a physical problem. The essential parallelism explored by this method makes it match with the powerful parallel computing resource GPU well to significantly improve the  the performance of GW data processing. After analyzing the execution and data access pattern of SPIIR, we design a novel method which can map the calculations of SPIIR onto the threads of GPU well. At the same time, we take advantage of the memory hierarchy and overlapping technology of GPU to improve the data access performance.  Our optimized implementation of SPIIR on GPU can achieve about 120 times speedup  compared with a single Intel i7 3770 CPU. The newly developed GPU SPIIR pipeline is capable of processing up to 21000 templates per GPU in real time. The  experimental results indicate that the full search for gravitational waves from compact binary coalescence in real time is possible with a fairly small number of desktop GPUs for the advanced LIGO detectors.}

\subsection{SPIIR method}
\label{sec:spiir}

 Summed Parallel Infinite Impulse Response (SPIIR) method \cite{shaun2} is a time-domain low latency algorithm for identifying the presence of gravitational waves produced by compact binary coalescence events in noisy detection data. Through the use  a parallel bank of single pole IIR filters, it is possible to approximate the SNR derived from the matched filter with greater than 99\% overlap. The main advantage of SPIIR method is that it operates completely in time domain, and in principle it has zero latency.

Another natural advantage of using a parallel bank of single-pole IIR filters is that they can be easily be executed in parallel. The essential parallel  feature of SPIIR makes it perfect to be implemented on modern multi-core or many-core processors such as GPU \cite{nickolls2010gpu,31}.

\subsection{GPU Computing}\label{gpu}
Graphics Processing Unit (GPU) is originally designed to address the demands of real-time high-resolution compute-intensive tasks. Since its first introduction in 1999 \cite{mcclanahan2010history}, GPU has been evolving fast that  many believe that GPU's power has not been fully exploited and its massive parallel nature can be leveraged for general purpose parallel computation. In late 2006, NVIDIA released its general purpose graphics processing unit (GPGPU) solution, the compute unified device architecture (CUDA) \cite{nvidia2011nvidia}, providing a general-purpose parallel computing platform and programming model  for its GPUs.

GPGPU technology evolves fast and it can provide very high performance mainly lies in that it can provide many parallel hardware threads. In hardware level, CUDA employs an SIMT (Single-Instruction, Multiple-Threads) architecture. A CUDA capable GPU consists of several SMs, each SM creates, manages, schedules, and executes threads in groups of 32 parallel threads called warps. A group of warps aggregate to a block, which is the basic and independent execution unit for an SM, all blocks launched by the user form a grid. This is the CUDA threads hierarchy and a typical CUDA application usually launches thousands of threads.

At the grid level, all the threads can share the global memory which is the largest but slowest memory of GPU. At the block level, all the threads of the same block can share the shared memory which is much faster than global memory. At the thread level, earch thread can only access its registers which are the fastest memory of GPU. But the latest warp-shuffle technology allows threads of the same warp to share their registers.

The key to achieve high performance on GPU is exploiting more parallelism in applications and accessing data in more efficiently way.

\subsection{Optimized GPU implementation of SPIIR}\label{optimization}
\begin{figure}[H]
\centering
\includegraphics[scale=1]{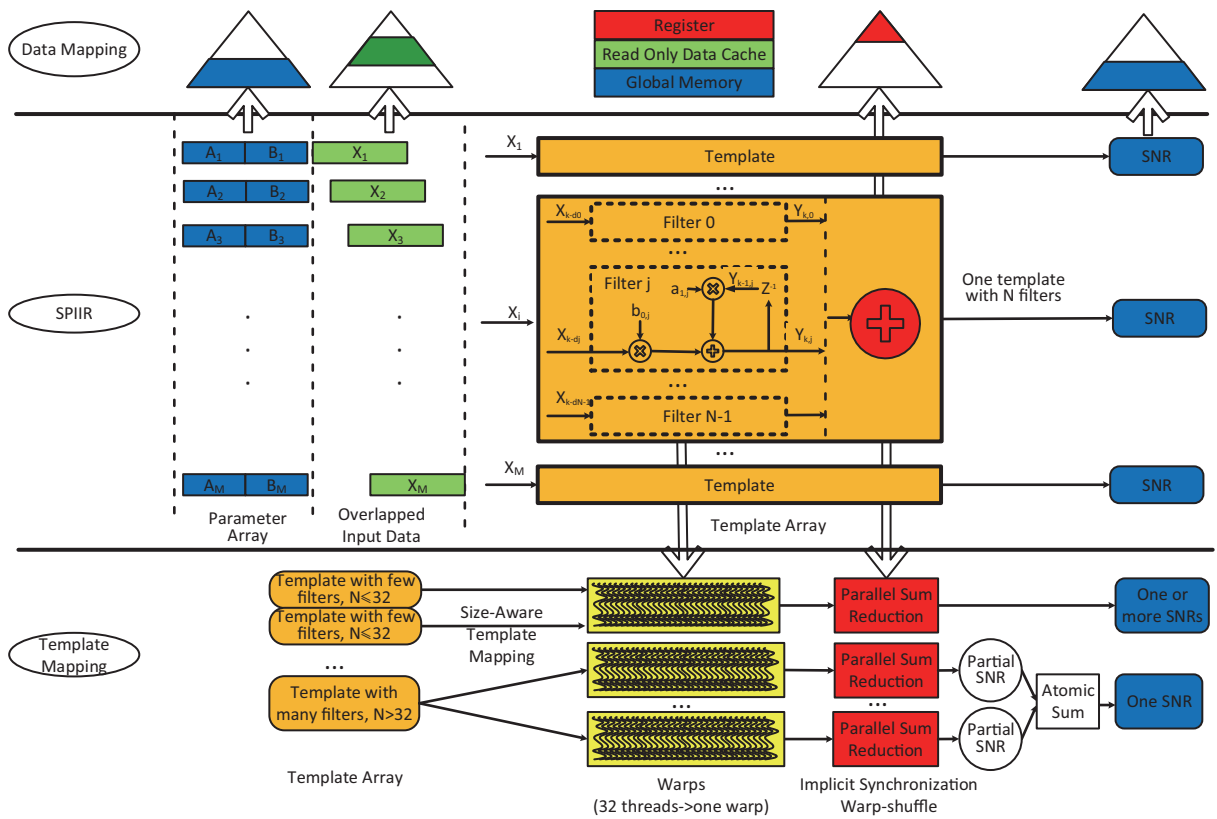}\vspace{-2mm}
		\captionof{figure}{(Color online) SPIIR method and its optimized implementation on GPU. The middle part shows how the SPIIR method calculates the SNRs from given parameter array and input data. The schematic diagram of one typical SPIIR template is described in detail. The up part shows we take advantage of the GPU's memory hierarchy to improve the performance of SPIIR data access. The bottom part shows how different SPIIR templates are implemented with CUDA on GPU to achieve low latency.}
		\label{fig:framework}
\end{figure}

The optimization method is very critical for GPU to achieve high performance\cite{opt_ppopp,liu2012gpu}.
Figure \ref{fig:framework} shows the framwork of the SPIIR method and its optimized implementation on Maxwell GPU. Our GPU implementation method can map the SPIIR filters onto the GPU thread hierarchy efficiently and take advantage of the GPU memory hierarchy to improve our method's performance.

For one typical SPIIR template with $N$ filters, to calculate the SNR at given time point, it will need the corresponding parameter array and the detection data within given time window. How to access those data efficiently is one important aspect to achieve low latency. Because the parameter array only need once to calculate one SNR, we organize them into aligned and continuous way to take advantage of coalesed memory access to load them from global memory with as few memory transactions as possible.

Because the time window of different  filters in one template are often overlapped, we take advantage of the read only data cache of Maxwell GPU to reuse the data for different filters. In this way, the threads of the same block can share the input data in read only data cache. Loading data from cache is much faster than from global memory, so reusing the data in cache can significantly improve the perpormance of data access.

We put the output data of each filter which is the input data in the next iteration into register to significantly reduce the data access delay. The SNR of one template needs the outputs of all its filters. We take advantage of the warf-shuffle feature of Maxwell GPU to enable different threads of the same warp to share the register data to improve the data access performance.

How SPIIR input and output data are organized and mapped onto the GPU memory hierarchy is shown in the upper part of Figure \ref{fig:framework}.

We design a size-aware template-warp mapping mechanism to efficiently map all  the SPIIR filters onto GPU thread hierarchy. Our mechanism can not only avoid branch execution paths in one warp which will significantly reduce the parallel performance of GPU, but also improve the utilization of GPU resources.

For a given template, if it has more than 32 filters, it will be mapped onto a CUDA block. Otherwise, it will be mapped onto a CUDA warp. In this way, we can avoid idle threads efficiently when a template has few filters.

We select suitable size of one CUDA block dynamically based on the template size and the number of registers of one thread to achieve approximate 100\% occupancy of GPU resources.

To calculate the SNR of a given template with $N$ filters, we need to sum all the output of different fileters in the template. Parallel sum reduction can reduce the number of sum operations from $(N-1)$ to  $\log_2N$ with $N$ parallel threads. On one side, the parallel method is helpful to improve perforamce. On the other side, it will also introduce synchronization operations among the parallel threads which is harmful for performance. Fortunately, the synchronization operation among all threads of the same warp can be guaranteed by hardware implicitely. So no explicit synchronization is needed. We take advantage of implicit synchronization feature of a CUDA warp to improve the performance.

The atomic operation is often expensive for old verion GPUs, but the latest GPU improve the performance of atomic operation significantly. So when we have to sum the partial SNRs from different warps, we can use atomic operation to calculate the result easily with very low cost.

How SPIIR filters are executed in the GPU thread hierarchy is shown in the bottom part of Figure \ref{fig:framework}. In the middle, a typical SPIIR template is given.

\subsection{Experimental results}

Our experiments are conducted on a machine with Intel \emph{Core} i7-3770 3.40 GH CPU and  8 GB DDR3 memory. The operation system is Fedora 20 64-bit and the host compiler is gcc 4.8.3. This machine is equipped with  NVIDIA \emph{GeForce} GTX 980 GPU with  CUDA Version  6.5.

Experimental results show that for the real pipeline data test, our GPU accelerated method is capable of handling 21000 templates in real time (one second). Compared with the CPU implementation, our GPU accelarated method can achieve more than 120X speedup.

We also check the presion of our GPU implementation and find that the maximum error is less than 0.1\% which is enough for the SPIIR method.

\subsection{Conclusion}
This work shows that optimizing design is very critical for us to achieve high performance on GPU. The exciting performance lies in the following factors: (1) the SPIIR method is parallel in essential so it is easy to explore fine-grain parallelism  and implement the method on GPU; (2) the latest GPU hardware provides additional computing power; (3) we develop a framework which can match the algorithm with the hardware features well.

\section{GPU acceleration revisited in $\bm\chi^{\bf 2}$ discriminator
  computation in compact binary coalescence searches}
  \emph{We demonstrate the speed-up that can be achieved by computing
the $\chi^2$ discriminator in compact binary coalescence (CBC)
searches on Graphics Processing Units (GPUs) rather than Central
Processing Units (CPUs). One of the most demanding gravitational-wave searches in
high-performance and high-throughput computing is that of CBC signals in multiple detectors. This is especially
true for those CBCs that have the potential to produce an electromagnetic (EM)
counterpart. Such a binary system will involve at least one neutron
star. The second object can be another neutron star or a relatively small
stellar-mass black hole. These systems not only have long-duration
signals but can require a large parameter space to define
them. Multiple detectors are needed to localize them in the sky for
ease of EM follow-ups. Finally, since the EM counterparts may fade
quickly, the searches need to be near-real-time
so as to enhance the chances of GW-EM coincidences.
These various factors require that the computing involved in analyzing
the GW data to find them be quick and inexpensive. We corroborate past
findings, but now with more modern~ hardware,  ~
that GPUs ~can provide~
significant }

\end{multicols}

\begin{multicols}{2}
\renewcommand{\baselinestretch}{1.08} \baselineskip 12.2pt\parindent=10.8pt
\renewcommand{\thefootnote}

\noindent   \emph{speed-ups in the computation of the
the $\chi^2$ discriminator in CBC searches, in comparison to not just single CPUs but also multiple MPI processes running on a CPU cluster.}

\subsection{Introduction}\label{introduction}

The Laser Interferometer Gravitational-wave Observatory (LIGO) and the
Virgo detector have shown that large scale
interferometers can be used to search for gravitational waves (GWs)
that are expected to change their arm-lengths at sub-nuclear scales.
With the next generation of gravitational wave interferometers,
with roughly ten times increased strain sensitivity, coming online soon the projected
event rate will increase multi-fold relative to what was expected in
the initial versions. These interferometers include the Advanced
LIGO \cite{AdLigoUrl} and the Advanced Virgo \cite{Virgo:2009} detectors,
the Japanese detector KAGRA \cite{Kuroda2010}, and possibly a LIGO detector in India.

One of the primary sources of GWs that these detectors will target is
the coalescence of a binary composed of neutron stars (NS) or black
holes (BH). Compact binary coalescences (CBCs) are the most promising
sources for detection with GW interferometric detectors. Not only are the estimated rates of CBCs in the search volumes of advanced detectors appreciable
but also their signals are theoretically well modeled.
The coalescences of compact binaries involving NSs (heretofore termed as CBCNSs) are especially attractive owing to the strong possibility of their being associated with electromagnetic (EM) counterparts. Unfortunately, the estimated rate of coincidences of GW detection of CBCs and their EM counterparts is not projected to be very high with networks of limited number of detectors.
Nonetheless, since coincident observations can teach us much more
about these sources, their origins, and their environments than what
any individual channel of observation can, it is all the more
important to explore ways of increasing the number of such
coincidences.

The primary objective of this paper is to revisit a solution
that was proposed by Chung et al. in
ref.~\cite{Chung:2009yb} for speeding up the computation of the
the signal-glitch $\chi^2$ discriminator, as defined by Allen in ref.~\cite{Allen:2004gu},
on Graphics Processing Units (GPUs).
This is just one of several algorithmic and hardware optimization
ideas that will be required to attain real-time implementation of CBC
searches, desirably in the data of multiple detectors.
A network of GW detectors is essential for localizing CBC sources in
the sky, so that EM observatories and particle detectors can look for their
counterparts in non-GW windows of Astronomy. However, since some of
these counterparts (e.g., kilonovae and X-ray or optical afterglows)
may not last very long after the GW trigger, it is crucial to be able
to search for GW candidates in the data of multiple detectors quickly
enough.
Indeed, the PyCBC software package (https://www.lsc-group. phys.uwm.edu/daswg/projects/pycbc.html) allows users to run various analyses in a CBC search pipeline on CPUs and GPUs in a transparent way. As observed in ref. \cite{31} computation in a CBC search is dominated by Fourier transforms, primarily in computing the signal-to-noise ratio and the $\chi^2$ discriminator for CBC waveforms drawn from a large template bank. In this work, we study the computational speed-up achieved from running one specific analysis, namely, the $\chi^2$ discriminator, on GPUs compared to CPUs. A clear drawback of our study stems from the fact that here we compute this discriminator not as a part of a CBC search pipeline but in isolation. Therefore, the actual speed-up realizable can be less owing to overheads arising from the integration of this analysis with the rest of the pipeline (which may involve slow processes, e.g., additional transfers of data from host to device across a PCIe bus). In that light, the utility of our study is that it explores the limit on the speed-up that may be achievable for this analysis when it gets run on GPUs. That, in turn, can influence hardware choices for future GW data analysis clusters.

Since the work of Chung et al. \cite{Chung:2009yb},
which is of 2010 vintage, several significant developements have taken
place. The list includes an increase in the total number of processing
elements (namely, threads), clock speed, and memory on
GPUs$^{1)}$\footnote{1) http://www.karlrupp.net/2013/06/cpu-gpu-and-mic-hardware-characteristics-over-time}. Further
progress has also taken place in the Nvidia CUDA framework in the form of additional libraries and algorithmic improvements in existing libraries$^{2)}$\footnote{2) {http://developer.download.nvidia.com/compute/cuda/6\_5/rel/docs/CUDA\_Toolkit\_Release\_Notes.pdf,} CUDA Release Notes for versions since 2010 are available online}.
In light of these developments, we are able to benefit from the following
improvements:

1. In ref.~\cite{Chung:2009yb}, the $\chi^2$ discriminator was
  tested on a 2.5 GHz Intel Q9300 CPU and an Nvidia GeForce 8800 Ultra GPU. This paper uses more modern, albeit not the most modern, hardware.

2.  While performance comparisons between an application
  parallelized to run on multiple threads on a GPU, on one hand, and a single-process, single-threaded
  version of the same application running on a CPU, on the other hand, can yield very large speed
  ups, in favor of the former, nonetheless this is not a fair comparison, some may
  argue. Therefore, we perform comparisons of the former with  the
  same application running on a {\em cluster} of CPUs, which may be termed
  somewhat more fair.

3. We use Advanced LIGO (as opposed to Initial LIGO) noise PSD for
  computing the $\chi^2$ discriminator, and the
  simulated signals and templates employed have a low-frequency cut-off of
  20 Hz. The CBCs studied have their component mass ranges, $m_1 \in
  [1.0, 3.0]$ M$_\odot$ and $m_2 \in
  [3.0, 20.0]$ M$_\odot$ (although it is more likely that to produce an
  EM counterpart $m_2$ should have an upper limit closer to $10.0$ M$_\odot$).

4.  This work uses up to 8000 GW templates, which is 8 times more than that used in ref.~\cite{Chung:2009yb}, that are selected from a
  much larger representative bank (with $\sim$50000 templates) of
  non-spinning NS-NS and NS-BH waveforms.

\subsection{Testing infrastructure}

The primary testing machine had two ``Intel(R) Xeon(R) CPU E5\-2650
$@$ 2.00 GHz'' processors, each with 8 cores. Each core is capable of
a parallel execution of 2 threads, thereby, leading to a possible $2\times 8
\times 2 = 32$ parallel threads. However, the CPUs on this machine
were used to run in a ``single-process, single-thread'' mode
alone. The same machine also had 128 GB RAM
and an Nvidia Tesla K20c GPU. The detailed configuration of the GPU card can be obtained from the Nvidia website.

Additional testing was done on the {\em Perseus} computing cluster,
whose configuration is described in Table 1. A salient point to consider is that
due to the lack of immediate availability of GPU clusters and very
large CPU clusters testing was restricted to a single run. In a future
work we plan to use multiple runs and use their average execution time
for comparisons.

\subsection{Parallelization strategy}
\label{section:parallelization:strategy}

We limited our $\chi^2$ computations to 8000 templates, ~of~ non-

\vspace*{3mm}
\begin{tablehere}
\caption{The specifications of the compute nodes of the Perseus Computing Cluster at the Inter-University Centre for Astronomy and Astrophysics, Pune, India are described here} \vspace{-2mm}\footnotesize
\begin{center} \doublerulesep 0.1pt \tabcolsep 9.7pt
\begin{tabular}{cl}
\toprule
Number of compute nodes& Configuration on each compute node\\
\hline
 & Two Intel(R) Xeon(R)\\
& CPU E5-2670 2.60 GHz,\\
94& with 8 cores each.\\
& 128 GB RAM. QDR Infiniband\\
& interconnect.\\
\bottomrule
\end{tabular}
\end{center}
\end{tablehere}

\begin{figure}[H]
\centering
\includegraphics[scale=1]{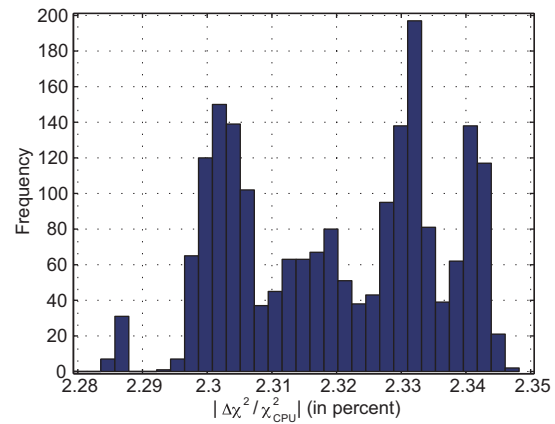}
\caption{(Color online) Differences in the $\chi^{2}$ values between the CPU and GPU
  runs for non-spinning CBC templates.}
\label{fig:diff:chi:sq:cpu:gpu:non:spin}
\end{figure}

\noindent spinning CBC signals,
because they sufficed to extract a large enough performance gap
between GPUs and CPUs. Their single-process and single-thread CPU
implementation is straightforward~\cite{Chung:2009yb}. For the GPU parallelization,
we adopted the simple strategy of computing the $\chi^2$ for each
template on a separate thread.
This is a viable strategy because (a) the computation for one template is
independent of that for the other templates$^{3)}$\footnote{3) This is also
  referred to in computer science parlance as an \textit{embarrassingly
    parallel problem}} and (b) the total amount of memory required is
less than 70 MB.

Another aspect of the GPU parallelization, specifically in the case of
the Nvidia GPUs, concerns the CUDA framework. It specifies a thread as
the principal unit for achieving computation. However, groups of
collaborating threads, or \textit{warps}, as they are known in the
CUDA framework, are logical parts of a ``block''. Each block is part
of a grid, which is set up and launched with a specified number of
blocks, and threads per block, in order to execute a kernel. Any
parallelization strategy on an Nvidia GPU must be mapped onto this logical arrangement.

\subsection{Results}

\subsubsection{Correctness}

The foremost result to be determined is how accurately the $\chi^2$
values are computed on the GPUs. To test this we compare them with
values obtained on CPUs for each of the same CBC mass pairs for the
non-spinning binaries. The results are shown in
Figure~\ref{fig:diff:chi:sq:cpu:gpu:non:spin}, which shows that
the fractional differences in the $\chi^{2}$ values
computed by the GPU and CPU are within 3\%, i.e., within acceptable limits.

\subsubsection{Performance results}

\begin{figure}[H]
\centering
\includegraphics[scale=1]{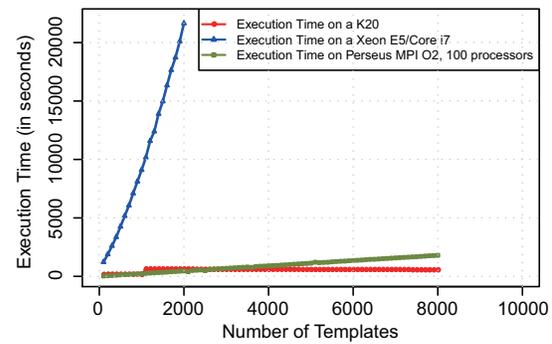}\vspace{-2mm}
\caption{(Color online) Comparison of execution times on various devices.\hspace{16mm}}
\label{fig:K20:MPI:O2:perseus:serial:xeon:e2650}
\end{figure}

The CPU results, while having the execution-time scaling linearly with
the problem size (which in our case is the number of templates), have
taken several hours for certain problem sizes, as shown in Figure~\ref{fig:K20:MPI:O2:perseus:serial:xeon:e2650}. In order to save time,
the tests for the single-process, single-threaded CPU were done
starting with a problem size of 100 templates, in steps of 100, and up
to 2000 templates.

\begin{figure}[H]
\centering
\includegraphics[scale=1]{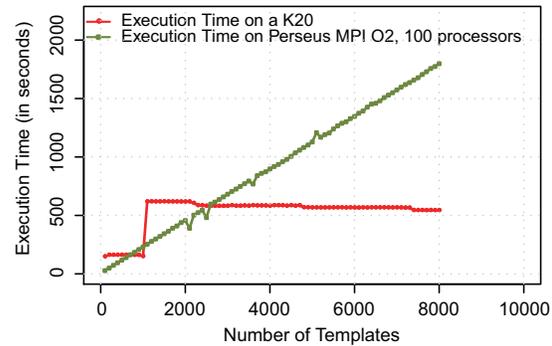}\vspace{-2mm}
\caption{(Color online) Comparison of execution time on the K20 and that on the
  Perseus computing cluster with MPI O2 and 100 parallel processors.}
\label{fig:K20:MPI:O2:perseus}
\end{figure}

\begin{figure}[H]
\centering
\includegraphics[scale=1]{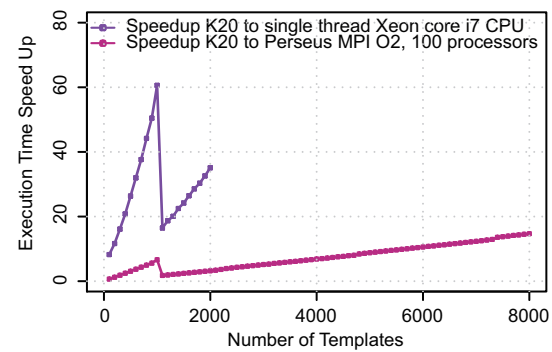}\vspace{-2mm}
\caption{(Color online) Execution time speed-ups on various devices.\hspace{20mm}}
\label{fig:speedups:K20:serial:xeon:CPU:E2650:MPI:O2:perseus}
\end{figure}

It is also seen in Figure~\ref{fig:K20:MPI:O2:perseus} that the GPU and MPI, with an
O2 optimization, and starting with a problem size of 100, in steps of
100, and up to 8000 templates for the non-spinning binaries, both finished
significantly quicker than the CPU,

As shown in Figure~\ref{fig:speedups:K20:serial:xeon:CPU:E2650:MPI:O2:perseus} the K20
predictably outperforms the single-process, single-threaded CPU by at
least a factor of $\sim$10 and as much as a factor of $\sim$60. What
is also interesting is that the K20 beats 100 parallel MPI
processes on Perseus.
We also find that the speedup against the 100 parallel MPI processes is
close to linear, albeit with a few inflection points.
At these points the performance of the execution time degrades for 1000 templates in the reference code since it uses the number of threads~ per~ block~ in ~ multiples~ of ~10,

\noindent which is sub-optimal. The Nvidia documentation recommends keeping the
number of threads per block to multiples of the warp size, which is 32. This is a potential future optimization of the code.

\subsection{Discussion}\label{sec:summary}

There is a need for exploring additional aspects of low-latency
searches than what was done here. First, in addition to providing computational
speed-ups for certain applications, GPUs can also be more efficient in
power consumption {\it vis \'{a} vis} CPUs. Drawing up a fair comparison of
power requirements for any application of interest, however, is somewhat
non-trivial. Second, CBCs with EM counterparts may not be limited to
binary neutron-star systems. Indeed, NS-BH binaries may also produce EM
counterparts. If so, non-spinning templates will not suffice in
ensuring the detection of GWs from such binaries. This is because back holes, more
than neutron stars, can retain large spins all the way up to
their GW emission driven merger with their binary
companions. Therefore, matched-filtering searches of GW signals from
NSBH systems must involve templates parameterized not just by the
component masses but also by an additional set of parameters that
describe the black hole spin. The resulting template bank can be orders of
magnitude larger than a non-spinning template bank (of the type
employed in this work).
Searching CBC signals with these templates may also benefit from the
use of GPUs in both computational speed up and more efficient power
usage, compared to CPUs.
Third, there are also algorithmic tricks that can be explored to speed up computation; see, e.g., ref. \cite{113} for one such insight in regards to the $\chi^2$ discriminator.
We plan to explore these ideas in a future
work.

\section{Multimessenger astronomy}
\emph{In this section we provide a short overview of the scope and strong future potential of a multi-messenger approach to gravitational-wave astronomy, that seeks to optimally combine gravitational wave and electromagnetic observations.  We highlight the importance of a multi-messenger approach for detecting gravitational wave sources, and also describe some ways in which joint gravitational wave and electromagnetic observations can improve the estimation of source parameters and better inform our understanding of the sources themselves---thus enhancing their potential as probes of astrophysics and cosmology.}

\subsection{Introduction}\label{sec:intro}%
As we mark the centenary of General Relativity, the rapidly emerging field of gravitational wave astronomy stands on the threshold of a new era, with the global network of so-called `second generation' ground-based interferometers preparing to begin operations \cite{Virgo:2009,Harry:2010}.  The approximately ten-fold improvement in sensitivity of these detectors compared with the first generation LIGO and Virgo detectors \cite{2015CQGra..32g4001T} means that the first direct detections of gravitational waves (GWs) are eagerly anticipated within the next few years---with the most likely sources expected to be the inspiral and merger of compact binary systems:  neutron star-neutron star (NS-NS) binaries; black hole-black hole (BH-BH) binaries or NS-BH binaries.   The challenges associated with detecting such sources, and exploiting them as probes of astrophysics and cosmology, have attracted considerable effort in recent years---including thorough investigations of the GW detection rates that can be expected \cite{2Abadie}, the most likely electromagnetic (EM) counterparts that will accompany them \cite{2012ApJ...746...48M} and the efficiency with which those EM counterparts could be detected \cite{2013ApJ...767..124N,2015arXiv150307869C}.  What is clear from this work is that significant benefits will derive from {\em joint\/} observations of both GW signals and their EM counterparts---giving rise to a new observational field that has been termed `multi-messenger' astronomy.

The benefits of a multi-messenger approach are manifest for both the detection and characterisation of GW sources.  Perhaps the most immediate benefit is a precise identification of the source sky location---alleviating the relatively poor sky localisation that is possible from ground-based GW observations alone \cite{Wen2008JPhCS.122a2038W,Wen2010PhRvD..81h2001W,2011ApJ...739...99N,Schutz2011CQGra..28l5023S,2011CQGra..28j5021F,2014PhRvD..89h4060S}.  Another clear benefit derives from the potential identification of an EM `trigger', or associated emission, that may streamline and enhance GW searches by reducing the number of free parameters, or at least significantly restrict the range of their values that must be explored \cite{2004CQGra..21S.765M,2008ApJ...683L..45A}.  Equally, a multi-messenger approach may involve using GW `triggers' to prompt targetted EM searches for the counterpart of the GW source \cite{2012A&A...541A.155A,2014ApJS..211....7A}---a successful outcome of which would indeed both identify a precise sky location and permit follow-up observation of the source's host galaxy,  allowing for example measurement of its redshift.  More generally, joint GW-EM observations may in turn lead to a better and more robust inference of the source parameters, thus opening the way to using GW sources more effectively as astrophysical probes.

There are very significant observational and computational challenges associated with these multi-messenger approaches, however.  From a GW point of view, there is the challenge of carrying out a sufficiently rapid real-time analysis of the interferometer data---thus allowing meaningful information to be supplied to the EM community quickly enough to permit follow-up observations of any prompt EM emission associated with the GW event \cite{2012A&A...541A.155A,2012A&A...539A.124L,2012ApJS..203...28E,2014ApJS..211....7A}.   To act upon the results of such an analysis then presents major challenges from an EM point of view:  e.g. understanding what {\em are\/} the EM counterparts themselves, and what are their signatures across the EM spectrum, and then searching efficiently for those signatures---the latter task rendered difficult in view of the poor sky localisation provided by the GW triggers, particularly during the first few years of operation of the second generation network \cite{2014ApJ...795..105S}.

In this paper we will consider some of these multi-messenger issues in more detail for the particular case of NS-NS or NS-BH compact binary coalescences (CBCs). We will very briefly summarise our current understanding of the expected EM counterparts of these sources and highlight some of the observational issues this raises for their detection through rapid follow-up observations across the EM spectrum.  We will also briefly discuss how GW searches can provide suitable triggers to the EM community on an appropriately short timescale.  Finally, we consider the astrophysical potential for multi-messenger observations of CBCs---highlighting one specific cosmological question that may be addressed in the fairly near future with these data.

Of course a multi-messenger approach is potentially crucial for {\em all\/} GW sources---not just CBCs---and may become increasingly important as the `detection era' of gravitational-wave astronomy unfolds.  For example, looking ahead to possible third generation ground-based detectors such as the proposed European Einstein Telescope \cite{2010CQGra..27s4002P} there are excellent prospects for multi-messenger science (involving not just GW and EM observations but also cosmic rays and neutrinos, see for example ref. \cite{2014PhRvD..90j2002A}) with a wide range of astrophysical sources.  The interested reader can find more details in e.g. refs. \cite{2011GReGr..43..437C,2013CQGra..30s3002A} but we do not consider the broader context for future multi-messenger astronomy any further in this paper.

\vspace*{-1mm}
\subsection{Joint detection of GW and EM emission:  constraints from the event timescale}\vspace{-1mm}

To make best use of a multi-messenger approach to observing

\noindent GW sources requires both a solid theoretical understanding of the signatures produced by the EM counterpart of the source, together with the ability to observe these signatures straightforwardly.  Consider the example of a NS-NS or NS-BH CBC merger event, for which the likely counterpart is thought to be a short-duration gamma ray burst (sGRB) \cite{1986ApJ...308L..43P,2005Natur.438..988B}.  Metzger \& Berger \cite{2012ApJ...746...48M} have addressed in detail the question of what might be the most promising electromagnetic counterpart of an sGRB, and they identify four key features (which they term `cardinal virtues') that such a counterpart should possess.  It should:

1. be detectable with present or upcoming telescope facilities, provided a reasonable allocation of resources;

2. accompany a high fraction of GW events;

3. be unambiguously identifiable, such that it can be distinguished from other astrophysical transients;

4. allow for a determination of $\sim$arcsecond sky position.

We will return to virtues 2, 3 and 4 in the next section, but the physical nature of a sGRB event makes satisfying virtue 1 already a challenge---particularly when one is considering EM searches prompted by a GW trigger---due to the very short timescales involved.   According to the standard picture, in the first one or two seconds  immediately after the merger the sGRB results from rapid accretion onto a centrifugally supported disk that surrounds the merged object; this accretion powers a highly relativistic jet that, due to beaming, is only seen as a sGRB within a narrow half-opening angle of a few degrees \cite{2006ApJ...653..468B,2006ApJ...650..261S}.  Interaction of the jet with the interstellar medium produces prompt afterglow emission that is visible in X-rays on a timescale of seconds to minutes after the merger, and in the optical on a timescale of hours to days---but again only within a narrow solid angle around the jet.  On the other hand it is believed that the event will also emit fainter isotropic radiation, in the optical and infra-red, on a timescale of days after the merger; this emission is caused by the decay of  heavy elements synthesised by the merger ejecta as it interacts with the surrounding environment, and is referred to as a `kilonova' \cite{1998ApJ...507L..59L,2005astro.ph.10256K,2010MNRAS.406.2650M,2015MNRAS.450.1777K}. We will see in the next section that kilonovae may be key to the successful identification of an EM counterpart for many CBC events.

Thus, as is summarised in e.g. refs. \cite{2014ApJ...795..105S,2015arXiv150803634S},  the nature of the EM counterparts of a NS-NS or NS-BH merger, and their rapid evolution in time, requires that the GW trigger be generated as quickly as possible---and ideally on a timescale of seconds to (at most) hours---particularly if the event afterglow is to be observed in X-rays.

Such a {\em low latency\/}, template-based data analysis pipeline has been developed for searching for CBC signals  with the ground-based interferometer network---and the need for a very rapid trigger generation is one of several key factors driving this search. (For more details see ref. \cite{2012ApJ...748..136C}).  In the sixth LIGO science run and third Virgo science run,  the inspiral search pipeline Multi-Band Template Analysis (MBTA) \cite{2008CQGra..25d5001B} was able to achieve trigger-generation latencies of 2--5 min (See e.g. ref. \cite{2012A&A...541A.155A}).  Interesting co-incident triggers identified by the MBTA were submitted to the Gravitational-wave Candidate Event Database (GraCEDb)$^{4)}$ \footnote{4) https://gracedb.ligo.org/}.  However, with the advanced detector network, given that sources will spend up to 10 times longer within the detection band, the  low-latency trigger generation of inspiral signals will become even more challenging.

The  slow evolution in frequency of an inspiral signal allows one to reduce the sampling rate of both the data and templates.   The  highly similarity between neighbouring templates in the filter banks allows one effectively to reduce the number of filters or to use  so-called `parallel  infinite impulse response'  filters.   We refer the reader to the LLOID (Low Latency Online Inspiral Detection) \cite{2012ApJ...748..136C} and SPIIR (Summed Parallel Infinite Impulse Response) \cite{2012PhRvD..85j2002L,shaun2,2012CQGra..29w5018L} pipelines for details of the implementation of these efforts.

\subsection{Joint detection of GW and EM emission:  identifying the source location}

From analysis of the GW interferometer network data alone, an approximate sky position for a CBC source can be obtained by exploiting the difference in arrival times of the GW signals at the different detector locations \cite{2013arXiv1304.0670L}.  In the first few years of operation of the advanced detectors the typical `error box' derived in this manner will, unfortunately, have an area of several hundred square degrees \cite{2014ApJ...795..105S}.  The situation will improve substantially in later years as more interferometers are added to the global network, and by the early 2020s---with the anticipated inclusion of the proposed LIGO India detector---almost $50\%$ of binary NS merger systems detected by the network are expected to be localised to within 20 square degrees on the sky.  Looking even further ahead, recent work \cite{2013CQGra..30o5004R,2015CQGra..32j5010H} investigating the optimal sites of future, third generation, detectors has applied various metrics to define and quantify the performance of a given network configuration---including the ability of the network to localise the position of a typical CBC source on the sky.  However, a fundamental limitation is set by the scale of the Earth, and the maximum difference which this imposes on the time of arrival of GWs at well separated detectors.  Thus, for {\em any\/} ground-based detector network, present or planned, we can expect that GW data alone will locate many sources to a precision of only tens or even hundreds of square degrees.  Clearly, then, identifying a unique EM counterpart would be a crucial benefit of a multi-messenger approach as it can provide a much more precise sky location for the GW source.

So how might such an EM counterpart be identified, given such relatively poor sky localisation information, in the context of the `cardinal virtues' highlighted above?  As discussed in ref. \cite{2012ApJ...746...48M}, sGRBs are not in themselves likely candidates since the beaming of their relativistic jets will mean that most merger events will not be seen as a sGRB at the Earth---even if a sGRB does indeed `accompany a high fraction of events'.  Similarly the prompt X-ray and optical afterglow emission associated with the sGRB may also not be seen at the Earth---although if such an afterglow {\em were\/} observed through rapid follow-up observations with a high-energy burst monitor satellite such as Swift, then this can in principle determine a very precise ($\sim 10$ arcseconds) sky location for the GW source \cite{2012ApJS..203...28E}.  As concluded in ref. \cite{2012ApJ...746...48M}, therefore, kilonovae are perhaps the most promising candidates for the EM counterpart of a NS-NS or NS-BH merger event since their isotropic emission gives them the potential to satisfy all four cardinal virtues.  On the other hand their comparative faintness and their limited duration still suggests that their detection will be very challenging, involving rapid, wide-field EM follow-up campaigns. In fact only three candidate kilonovae events have been observed \cite{2013Natur.500..547T,2015NatCo...6E7323Y,81}.

Whatever the nature of its EM counterpart, a key step towards its study is the (hopefully  unique) identification of the host galaxy of the GW source.  To this end we can develop a general framework for combining GW and EM information under the straightforward assumption that there are common parameters---such as sky location and distance---among the observations made of a GW source, its EM counterpart  and their host galaxy.  Such a multi-messenger approach has already proven fruitful even in the {\em absence\/} of any observed GWs.  For example, searches for GWs associated with the gamma-ray bursts  GRB070201 \cite{2008ApJ...681.1419A} and GRB051103 \cite{2012ApJ...755....2A}  have ruled out the possibility that their progenitors were binary neutron stars (NS)  sources in Andromeda galaxy (M31) and M81, respectively.

One can consider the question, then, of how the detection efficiency of an EM counterpart might be improved by conducting wide-field EM follow-ups that make use of pre-existing galaxy catalogs.  This issue has been investigated in ref. \cite{2014ApJ...784....8H}; by taking into account the GW measurement error in both distance and sky location they estimated that an average of $\sim 500$ galaxies are located in a typical GW sky location error box for NS-NS mergers with Advanced LIGO ($\sim$20 deg$^2$), up to range of 200 Mpc. The authors then found that the use of a complete reference galaxy catalog could improve the probability of successful identification of the host galaxy by $\sim 10\%$--$300\%$ (depending on the telescope field of view) relative to follow-up strategies that do not utilize such a catalog.

Of course a more complete treatment ideally should involve a comprehensive end-to-end simulation that takes fully into account the characteristics of the global GW network, the spatial distribution and event rate of CBC events, the multi-wavelength light curves of their EM counterparts (together with the event rate and lightcurves of a range of possible `false positives' with which these counterparts might be confused, bearing in mind virtue 3 above) and the sensitivies and observing strategies adopted by the multi-wavelenth EM telescopes used to search for them.  Much work remains to be done in this area, both on more detailed theoretical modelling of the EM counterparts themselves (e.g. kilonovae) and on investigating the impact and efficacy of different follow-up observing strategies.  Some excellent progress has already been made, however, particularly in ref. \cite{2013ApJ...767..124N} where the authors point out that:

$\bullet$ dedicated 1 m class optical and near infra-red telescopes with very wide-field cameras are well-positioned for the challenge of finding EM counterparts

$\bullet$ a comprehensive catalog, out to $z$ $\sim$ 0.1, of foreground stellar sources, background active galactic nuclei and potential host galaxies in the local universe---and probing deeply (i.e. approximately 2 magnitudes deeper than the expected magnitudes of the EM counterparts)---is needed to assist identification of the EM counterparts amidst the larger numbers of false positives.

To date no such all sky galaxy catalog is available.  Among the many existing galaxy catalogs, the Gravitational Wave Galaxy Catalog \cite{White2011CQGra..28h5016W} (GWGC; (White et al. 2011)) has been specifically compiled for current follow-up searches of optical counterparts from GW triggers---although the authors acknowledged that the catalog is increasingly incomplete at larger distances. Nuttall $\&$ Sutton \cite{2010PhRvD..82j2002N} proposed a ranking statistic to identify the most likely GW host galaxy (drawn from the GWGC) based on galaxy distance and luminosity and the sky position error box. This ranking method has been adopted in the design of an EM follow-up pipeline  (such as ref. \cite{2013ApJS..209...24N})  and follow-up observations (such as refs. \cite{2014ApJS..211....7A,2012ApJ...759...22K,AndoRevModPhys.85.1401}).

 A novel Bayesian approach to identifying the GW host galaxy, incorporating astrophysical information has been proposed in ref. \cite{2014ApJ...795...43F}. A merit of this approach is that both the rank and the posterior probability of a galaxy hosting the GW source are estimated with the help of  GW-EM relation models.   Using a simulated population of CBCs the authors found that (i) about 8\%, 4\%, and 3\% of injections had 50\%, 90\%, and 99\% respectively of the probability to be included in the top 10 ranked galaxies in the GWGC, and (ii) the first ranked galaxy had a 50\% probability of being the true GW host galaxy in about 4\% of injections. These results are dominated by the GWGC distance cut of 100 Mpc, compared with the expected reach of $\sim$200 Mpc for Advanced LIGO and Advanced Virgo at design sensitivity.  The method could easily be extended and applied to deeper galaxy catalogs, however.

A comprehensive, complete  all sky catalog of {\em all\/} objects (galaxies, stellar sources and so on) may not be essential  for a single GW-EM joint observation, since the  GW sky map will only cover a small fraction  of the whole sky.   A ``galaxy survey on the fly" approach has been proposed to reduce the required effort and thus enhance the immediate availability of useful catalogs \cite{2015ApJ...801L...1B}; in this approach a rapid galaxy survey using 1--2 m class telescopes can efficiently catalog those galaxies covered by one GW detection volume within a short period of time.   This rapidly compiled galaxy catalog could then very quickly be provided to other telescopes, to aid further electromagnetic follow-up observations of e.g. kilonovae, as well as other sources.

Whatever detailed strategies are employed in the future, it is clear that to identify the common GW and EM source one needs multi-wavelength  observations capable of rapidly and efficiently targeting multiple objects.  A community of `multi-messenger' astronomers---including the LIGO and Virgo collaborations and a significant and growing number of telescopes across the world---is now being assembled for exactly this purpose$^{5)}$\footnote{5) https://gw-astronomy.org/wiki/viewauth/LV\_EM/}.

\subsection{Beyond detection}

The promise of multi-messenger astronomy---and in particular of joint GW and EM observations of CBCs and their EM counterparts---lies not just in enhancing the prospects for detecting these events but also in improving the estimation of their parameters, and ultimately their use as astrophysical and cosmological probes.

Consider again, for example, the case of a CBC event. In the absence of any spin, the inspiral phase of a CBC in a circular orbit has a gravitational waveform that depends on the following nine parameters: four intrinsic parameters---the two masses and two intrinsic constants of integration, which define the phase evolution of the waveform; five extrinsic parameters---luminosity distance, sky location, inclination angle (the angle between the orbital angular momentum and the line of sight) and GW polarization.  All nine parameters govern the amplitude of the signal that each GW detector should `see'.  So we see immediately that the identification of a unique EM counterpart will begin to reduce the dimensionality and complexity of this parameter space, providing tight constraints on the sky location and the luminosity distance (e.g. via a measured redshift for the host galaxy---although deriving a luminosity distance directly from this will be complicated by the effects of galaxy peculiar velocities).  Knowledge of the luminosity distance from an EM counterpart will also help to break parameter degneracies that may exist for the GW source, such as that between distance and inclination angle \cite{2010ApJ...725..496N}.

More generally, joint GW-EM observations should allow for better characterization of the signal progenitor, and a richer interpretation of the results of GW searches---offering, for example, greater insight at the population level into the relationship between the GW source, its EM counterpart(s) and their progenitor environment.  An excellent example of such an approach is the relationship between CBCs and sGRBs that we have discussed in detail in this paper, but as the detection era unfolds no doubt other similar examples will emerge.  The insights derived from these joint analyses may improve our understanding of the evolutionary processes that led to the GW event, and thus perhaps provide better prior information about the likely sites (e.g. in terms of host galaxy morphology, colour, metallicity etc) of future events.   Moreover, as our understanding of these deep, underlying astrophysical relationships improves, new data analysis methods will be built to explore them---combining not just GW and EM observations but also data from neutrino and cosmic-ray telescopes \cite{2013JCAP...06..008A,2014PhRvD..90j2002A}.

Underpinning this joint approach at all stages is the fundamental assumption that these multi-messenger phenomena all emanate from the same object, and obey physical relationships that share some common parameters.  Bayesian inference then provides a natural framework in which to incorporate multi-messenger astrophysical information and optimally estimate those parameters. The method introduced in refs. \cite{2014ApJ...795...43F,2015ASSP...40...35F} explores some specific examples of this framework in action; these include reducing the area of sky over which astronomical telescopes must search for an EM counterpart, improving the inference on the inclination angle of a NS-NS binary and estimating the luminosity of a sGRB.

What specific astrophysical questions will be tackled via a multi-messenger approach in the future?   We end by briefly considering one such example that is anticipated for the advanced detector era.

It was recognised nearly thirty years ago by Schutz \cite{Schutz1986Natur.323..310S} that CBCs have the potential to be precise luminosity distance indicators via measurement of the time-evolution of their amplitude, frequency and frequency derivative during the inspiral and merger phase.  This has given rise to the idea of so-called gravitational-wave {\em standard sirens\/}, by analogy with the {\em standard candles\/} of EM astronomy---although the CBCs do not require any assumption to be made about their intrinsic `luminosity' (which essentially depends on the masses of the binary stars) as this can be inferred directly from the observed data at the same time as the luminosity distance.  In more recent years there has, therefore, been much interest in the potential use of standard sirens as cosmological probes, via calibration of the luminosity distance redshift relation.

The expected reach of advanced detectors will be too shallow to permit exploration of dark energy models and the accelerated expansion of the Universe---although such models could certainly be investigated by third generation detectors such as the Einstein Telescope \cite{2010CQGra..27u5006S,2011PhRvD..83b3005Z} or spaceborne missions such as eLISA \cite{2005ApJ...629...15H,2007ApJ...668L.143D,2012CQGra..29l4016A}.  However, a realistic target for the upcoming global network of advanced detectors is measurement of the Hubble constant, $H_0$, using standard sirens.  Recently Nissanke et al. \cite{2013arXiv1307.2638N} have investigated the efficacy of such a measurement,  and conclude that a precision of about $1\%$ on $H_0$ is possible from observations of about 30 NS-NS mergers within a few hundred Mpc---using the anticipated future global intereferometer network that includes KAGRA and LIGO India.  Such a precise value would certainly be competitive with the EM results expected on a similar timescale from e.g. the James Webb Space Telescope \cite{2010ARA&A..48..673F}, and in any case would be an extremely useful adjoint to a purely EM determination of $H_0$ using the traditional cosmic distance ladder since it would be subject to a completely different set of systematic uncertainties.

A standard siren measurement of $H_0$ will present a major multi-messenger challenge for several reasons.  Firstly, to estimate the Hubble constant of course requires comparison of distance with redshift, and the latter will not generally be measurable from GW data alone (but see also below for discussion of some interesting alternative approaches). Indeed the luminosity distance estimates for the sirens will in any case be degenerate with redshift because of the mass-redshift degeneracy in post-Newtonian CBC waveforms.  By measuring the redshift of the siren's host galaxy the degeneracy can immediately be broken.  However, this measurement of course first requires the prompt observation of an EM counterpart and the unique identification of the host galaxy---steps which will be subject to all of the multi-messenger issues discussed in the previous section.

Notwithstanding these potential difficulties, and their resulting impact on the final error budget for $H_0$ (for which the the estimate of $1\%$ in ref. \cite{2013arXiv1307.2638N} may therefore be somewhat too optimistic), the prospect of a gravitationally-calibrated value of the Hubble constant is nevertheless extremely exciting---and is likely to be one of the main targets for gravitational-wave astronomy over the next decade.

It is interesting to note that some other approaches to breaking the mass-redshift degeneracy, and/or determining redshifts (and hence cosmological parameters) from GW data alone, have been proposed.  For example Taylor et al.  \cite{2012PhRvD..85b3535T} assume that there exists a universal (rest frame) mass distribution for NSs at different redshifts and by comparing the measured (redshifted) mass distribution of NSs with the local mass distribution show that one could infer statistically the redshifts of the sources and hence derive indirectly the value of $H_0$. Their results show that, in this way, second generation interferometers should be able to infer the Hubble constant with $\sim 10\%$ accuracy from about 100 events. Their analysis is extended in ref. \cite{2012PhRvD..86b3502T} to consider the cosmological potential of GW-only observations with third generation ground-based detectors.

In a similar manner, various authors \cite{2008PhRvD..77d3512M,2011ApJ...732...82P} have proposed that the identification of the host galaxy---and thus the determination of its redshift---may be carried out {\em statistically\/}, using prior information about the spatial distribution of the galaxies in the sky localisation error box provided by the GW data alone.  Using this formalism del Pozzo  \cite{2014JPhCS.484a2030D} suggests that the Hubble constant could be determined to a few percent from observations of about 50 sirens with the advanced detector network.

A third, highly promising, possibility for constraining cosmological models using GW observations alone has been proposed by Messenger \& Read \cite{2012PhRvL.108i1101M}, exploiting the effect of tidal deformations on NS-NS binary systems during the inspiral that provide additional contributions to the phase evolution of their gravitational waveforms. Recently their approach has been studied further, and is predicted to be capable of determining redshifts to a precision of 10$\%$--20$\%$ for GW sources in the local Universe observed with the Einstein Telescope \cite{2014PhRvX...4d1004M}.   The potential of this approach for determining cosmological parameters with Einstein Telescope observations has been further explored in ref. \cite{2015arXiv150606590D}.

While these various methods suggests the intriguing possibility, therefore, of measuring $H_0$---and indeed other cosmological parameters---without using {\em any\/} EM observations directly, the generally lower precision of these estimates does, nonetheless, underline that a multi-messenger approach will usually be more effective than a GW-only analysis.

\subsection{Summary}

In this paper we have discussed some of the advantages---both for the detection of GW sources and also for the estimation of their parameters and their astrophysical exploitation---of a multi-messenger approach that seeks to combine optimally GW and EM observations.  Focussing mainly on the inspiral and merger of NS-NS and NS-BH binaries, which are believed to be the progenitors of short duration GRBs, we have also highlighted some of the important observational challenges that need to be overcome in order that a multi-messenger approach may be fully exploited.  These challenges present significant logistical and computational constraints for the analysis of data from the ground-based network of advanced GW interferometers.  This is because of the very short timescale (from seconds to hours to days at most) associated with the EM counterparts of these events and the relatively poor sky localisation provided by the GW data alone---which will significantly complicate the search for a unique EM counterpart. Nevertheless substantial progress has already been made towards establishing a community of `multi-messenger' astronomers, working closely together towards the goal of making joint GW-EM observations, and there are excellent prospects for this emerging new field over the next decade.

Looking further ahead, the potential of multi-messenger astronomy would appear to grow even stronger with the possible advent of third generation ground-based interferometers such as the Einstein Telescope and the possible launch of a spaceborne GW detector such as eLISA.  Although we did not discuss these future missions and projects in this paper, they should offer exciting science possibilities such as probing the equation of state and internal structure of neutron stars, constraining models of core-collapse supernovae and mapping the detailed structure of spacetime around black holes.  As general relativity enters its second century, the prospects for testing Einstein's theory under extreme cosmic conditions using multi-messenger data look very bright.

\vspace*{2mm}\Acknowledgements{\bahao This work was supported by the NRF from the Korean government (Grant No. 2006-00093852), the National Natural Science Foundation of
China (Grant Nos. 61440057, 61272087, 61363019, 61073008 and 11303009), Beijing Natural Science
Foundation (Grant Nos. 4082016 and 4122039), the Sci-Tech Interdisciplinary
Innovation and Cooperation Team Program of the Chinese Academy of
Sciences, the Specialized Research Fund for State Key Laboratories, National Science Foundation (Grant Nos. PHY-1206108 and PHY-1506497), the Perseus
  Computing Cluster at the Inter
University Centre for Astronomy and Astrophysics (IUCAA), Pune,
India. BOSE Sukanta thanks the Kavli Institute for
  Theoretical Physics China (KITPC), Beijing, and the organisers of
  the KITPC program on Next Detectors for Gravitational Wave Astronomy,
  for their hospitality.
  BOSE Sukanta, PANDEY Vihan and PHUKON Khun Sang wish to thank DAL CANTON Tito, GUPTA Anuradha, JAIN Pankaj for helpful discussions and S. PARSURAMAN and GAJBE Alpesh of the Tata Institute of Social Sciences for their invaluable support.
  KSP thanks IUCAA for its hospitality during his stay there. HENDRY Martin acknowledges the hospitality and financial support provided by the Kavli Institute for Theoretical Physics in Beijing.}

\end{multicols}

\end{document}